\documentstyle[epsfig]{europhys}

\def\etal{{\hbox{{\tenit\ et al.\/}\tenrm :\ }}}

\def\And{{\rm and\ }}

\def\drm{{\rm d}}

\newif\ifboo \boofalse

\def\Review#1{\boofalse{\it #1},}
\def\Name#1{{\sc #1},}
\def\Vol#1{\ifboo Vol. {\bf #1}\else{\bf #1}\fi}
\def\Year#1{\ifboo #1\else(#1)\fi}
\def\Book#1{\bootrue{\it #1},}
\def\Page#1{\ifboo {\rm p. #1}\else{\rm #1}\fi}

\newcommand{\gtrsim}{
\,\raisebox{0.35ex}{$>$}
\hspace{-1.7ex}\raisebox{-0.65ex}{$\sim$}\,
}

\newcommand{\lesssim}{
\,\raisebox{0.35ex}{$<$}
\hspace{-1.7ex}\raisebox{-0.65ex}{$\sim$}\,
}

\begin{document}
%
%
\euro{46}{4}{425-430}{1999}
\Date{15 May 1999}
\shorttitle{K. KLADKO \etal CUMULANT EXPANSION FOR SYSTEMS WITH LARGE SPINS }
\title{Cumulant expansion for systems with large spins}
\author{
K. Kladko\inst{1}
\footnote{cnls.lanl.gov/$\sim$kladko/; kladko@lanl.gov},
P. Fulde\inst{2},
\And D. A. Garanin\inst{2} 
\footnote{www.mpipks-dresden.mpg.de/$\sim$garanin/; 
garanin@mpipks-dresden.mpg.de
}
} 
\institute{
     \inst{1} Condensed Matter Physics Group and Center for Nonlinear Studies,\\  
Theoretical Division, Los Alamos National Laboratory,  \\
MS-B258, Los Alamos, New Mexico 87545, USA \\
     \inst{2} Max-Planck-Institut f\"ur Physik  komplexer Systeme,\\
N\"othnitzer Str. 38, D-01187 Dresden, Germany 
}
%
%
\rec{23 November 1998}{12 March 1999}
%
\pacs{
\Pacs{05}{30.-d}{Quantum statistical mechanics}
\Pacs{75}{10.Jm}{Quantized spin models}
\Pacs{75}{10.Hk}{Classical spin models}
}
\maketitle
\vspace{-1cm}
\begin{abstract}
A method is proposed for obtaining a systematic expansion of thermodynamic functions of spin
systems with large spin $S$ in powers of $1/S$.
It uses the cumulant technique and a coherent-state representation of 
the partition function ${\cal Z}$.
The expansion of ${\cal Z}$ in terms of cumulants yields an effective classical
Hamiltonian with temperature-dependent quantum corrections.
For the
Heisenberg quantum Hamiltonian, they have a  non-Heisenberg form.
The effective Hamiltonian can be solved by methods familiar for classical
systems.
\end{abstract}

\vspace{-1cm}

In recent years considerable interest in condensed matter 
physics has been focusing on the properties of new magnetic materials,
such as magnetic molecules with a  large effective spin $S$, or
intermetallic compounds containing magnetic layers or chains.
For the latter, most of the theoretical studies deal with  $S=1/2$ systems
which, in the one-dimensional case, allow for analytical solutions using 
special methods such as the Bethe Ansatz or Jordan-Wigner fermion representation of spins.
On the other hand, constructing expansions in powers
 of $1/S$
for systems with large values of S, which are not constrained to special
lattice geometries or to low temperatures, has not received sufficient attention.

The aim of  this paper is to show that quantum corrections to the 
classical thermodynamics of spin systems with large or moderate $S$
can be calculated to an 
arbitrary order in the quasiclassical parameter $1/S$. 
This is achieved by using a {\em cumulant expansion technique} and working
in the basis of {\em spin coherent states}.
The results which can be obtained by our method are valid for temperatures down to the quantum energy scale
$\tilde J/S$, where $\tilde J \equiv JS^2$ is the classical energy scale
and $J$ is the exchange interaction.
They are complementary to those following from the spin-wave theory
(SWT) based either on the Holstein-Primakoff or
path-integral formalisms, which works in the range $T \lesssim \tilde J$
(see, for example, ref.\ \cite{aue94book}).

Averages or matrix elements $\langle \ldots\rangle$ can be expressed through 
{\it cumulants} $\langle \ldots\rangle^c$ as follows
%
\begin{eqnarray}\label{AvrToCum}
&&
\langle  A \rangle=\langle  A \rangle^{c},
\qquad
\langle  A_1 A_2  \rangle=\langle A_1 A_2  \rangle^{c}
+
\langle  A_1 \rangle \langle A_2  \rangle, 
\\
&&
\langle  A_1 A_2 A_3 \rangle = \langle  A_1 A_2 A_3 \rangle^{c}
+\langle  A_1 \rangle \langle A_2 A_3 \rangle^{c}
+\langle  A_2 \rangle \langle A_1 A_3 \rangle^{c}
+\langle  A_3 \rangle \langle A_1 A_2 \rangle^{c}
+\langle  A_1 \rangle \langle  A_2 \rangle \langle  A_3 \rangle,\nonumber
\end{eqnarray}
etc., where $A_i$ are classical stochastic variables or
quantum-mechanical operators.
The averaging above is performed over a classical distribition function or weighed over
quantum states. 
A detailed discussion of cumulants, with an emphasis on quantum systems, can be found in refs.\
\cite{ful95book,klaful98}.
Cumulants can be obtained by differentiation of a generating function,
i.e., from 
%
\begin{equation}\label{AvrToCumLog}
\left.\langle A_{1}\ldots A_{N} \rangle^{c}= \frac{\partial}{\partial
\lambda_1 }\cdots \frac {\partial}{\partial
\lambda_N}~ 
\ln \langle  e^{\lambda_1 A_1} \ldots e^{\lambda_N A_N} 
 \rangle~\right|_{\lambda_{1}=\ldots=\lambda_{N}=0}, 
\end{equation}
in contrast to averages $\langle\ldots\rangle$, which are given by a similar
expression without logarithm.
By setting $A_i =A$, multiplying by $\lambda^n/n!$, and summing over
$n=1,\ldots,\infty$ one obtains the identity
%
\begin{equation}\label{ExpCumRel}
\ln\langle e^{\lambda A} \rangle = \langle e^{\lambda A} -1 \rangle^c. 
\end{equation}

Let us now point out a fundamental relation between cumulants and
a quasiclassical theory.
It is a {\em characteristic}  property of quantum mechanics, 
that the expectation value
of a product of two observables $A_1$ and $A_2$, evaluated with respect to some
quantum  state $|\psi\rangle$, is generally distinct from a product of the
expectation values of  $A_1$ and $A_2$, i.e., 
%
\begin{equation}\label{QuantumNonZero}
\langle A_1 A_2\rangle^c = \langle A_1 A_2\rangle 
- \langle A_1\rangle \langle A_2\rangle \ne 0,
\end{equation}
whereas equality is achieved in the classical limit.
Thus one  expects that the theory formulated in terms of cumulants will yield
a quasiclassical expansion.

To this end, we rewrite the partition function ${\cal Z}$ of the system in the basis of  
{\em spin  coherent states} $|{\bf n}\rangle$, i.e., maximum-spin states pointing in the direction of the
unit vector ${\bf n}$.
In this basis, the cumulants have a simple form, especially if the spin-operator components
$S_z$ (axis $z$ along ${\bf n}$) and $S_\pm=S_x\pm iS_y$ are used. 
First, a cumulant vanishes if it is  of the form $\langle \ldots S_+\rangle^c$,
$\langle \ldots S_+S_-\rangle^c$, $\langle \ldots S_z\rangle^c$, etc.
Non-vanishing cumulants are
%
\begin{eqnarray}\label{LowestCums}
&&
\langle S_+ S_z^n S_- \rangle ^c = 2S (-1)^n,
\nonumber\\
&&
\langle S_+ S_z^n S_+ S_z^m S_- S_z^{n'} S_- \rangle ^c = -4S (-1)^{n+n'} (-2)^m, \nonumber\\
&&
\langle S_+ S_+ S_+ S_- S_- S_-\rangle ^c 
= 3 \langle S_+ S_+ S_- S_+ S_- S_-\rangle ^c = 24S,
\end{eqnarray}
etc.
The above results can be obtained recurrently, using eq.\ (\ref{AvrToCum}) and commutation
relations $[S_z,S_-]=-S_-$ and $[S_+,S_-]=2S_z$, which remain valid inside  cumulants.
For scaled spins 
${\bf S}/S$, each nonvanishing cumulant containing $n$ spin operators scales as 
$1/S^{n-1}$.

In the overcomplete basis of spin coherent states, the unit operator is resolved as
%
\begin{equation}\label{ResUnity}
{\bf 1} = \frac{2S+1}{4 \pi} \int \drm{\bf n}\, |{\bf n}\rangle \langle{\bf n}|. 
\end{equation}
Thus, for a single-spin system, the trace of an operator $\hat A$ over any complete orthonormal 
basis $|m\rangle$ 
may be written in terms of spin coherent states as follows \cite{lie73}
%
\begin{equation}\label{TraceRewrite}
{\rm tr}\, \hat A = \sum_m \langle m| \hat A | m \rangle 
= \frac{2S+1}{4 \pi} \int \drm {\bf n} 
\sum_m \langle m|\hat A|{\bf n}\rangle \langle {\bf n}|m\rangle 
= \frac{2S+1}{4 \pi} \int \drm {\bf n}\, \langle {\bf n}| \hat A| {\bf n}\rangle.
\end{equation}
The partition function for a Hamiltonian $\hat H$ becomes
%
\begin{equation}\label{ZCohState}
{\cal Z} =  {\rm tr} \exp (- \beta \hat H) = 
\frac{2S+1}{4 \pi} \int \drm {\bf n}\, \langle {\bf n}| \exp (- \beta \hat H) |
{\bf n}\rangle, \qquad \beta \equiv \frac 1T. 
\end{equation}
The matrix element inside the integral can be written with the help of 
eq.\ (\ref{ExpCumRel}) as 
%
\begin{equation}\label{ExpCum}
 \langle {\bf n}| \exp (- \beta \hat H) |{\bf n}\rangle 
 = \exp[-\beta {\cal H}({\bf n},\beta)],
 \qquad \beta {\cal H}({\bf n},\beta) =
  \langle {\bf n}|1 - \exp (- \beta\hat H) |{\bf n}\rangle ^c ,
\end{equation}
where ${\cal H}({\bf n},\beta)$ can be considered as an effective quasiclassical
Hamiltonian.
In order to find an expansion of 
${\cal Z}$ in powers of $1/S$, we expand the expression for ${\cal H}$ 
 in a Taylor series:
%
\begin{equation}\label{HEffExp}
 {\cal H}({\bf n},\beta) =  
  \langle {\bf n}|\hat H|{\bf n}\rangle^c 
 - \frac{\beta}{2!} \langle {\bf n}| \hat H \hat H|{\bf n}\rangle^c 
 + \frac{\beta^2}{3!} \langle {\bf n}|\hat H \hat H \hat H|{\bf n}\rangle^c + \ldots
\end{equation}
Here the first term on the right-hand side is the classical energy of the spin.
As seen from eq.\ (\ref{LowestCums}), increasing powers of $1/S$ appear
in each order of this expansion.
If the quantum corrections in eq.\ (\ref{HEffExp}) are small, then in most cases
one can further expand $\exp(-\beta {\cal H})$ in powers of $1/S$.
In this case the {\em statistical weights} of different spin orientations are described by
a purely classical Hamiltonian, whereas quantum effects  manifest themselves in
corrections to the {\em density of states}.

As an example, let us consider the problem of a spin in a magnetic field.   
%
\begin{equation}\label{HamField}
\hat H = -  {\bf S}\cdot  {\bf H}.
\end{equation}  
For this model, the effective  Hamiltonian 
${\cal H}({\bf n},\beta)$ can be
computed analytically from eq.\ (\ref{ExpCum}), if one expresses the coherent states
in the basis of states
with definite projections of ${\bf S}$ on ${\bf H}$.
This yields
%
\begin{equation}\label{HamFieldHeffExact}
{\cal H} = -T \ln \sum_{m=-S}^S e^{m\beta H} \frac{ (2S)! }{ (S+m)! (S-m)! }
\left( \frac{ 1+\cos\theta }{ 2 } \right)^{S+m} 
\left( \frac{ 1-\cos\theta }{ 2 } \right)^{S-m},
\end{equation}  
where $\theta$ is the polar angle.
To obtain the cumulant expansion of ${\cal H}$,  it is convenient
to represent the spin operator ${\bf S}$ in eq.\ (\ref{HEffExp}) in the coordinate
system with the $z$ axis along the vector ${\bf n}$  
%
\begin{equation}\label{SpinCoord}
{\bf S} = {\bf n} S_z +  {\bf n}_+ S_+ +  {\bf n}_- S_-,
    \qquad {\bf n}_\pm \equiv ({\bf n}_x \mp i{\bf n}_y)/2,
\end{equation}  
where ${\bf n}_x$ and ${\bf n}_y$ are appropriate transverse basis vectors.
Then cumulant averages of different spin components are done with the use of 
eq.\ (\ref{LowestCums}).
To order $1/S^2$  one obtains
%
\begin{equation}\label{HamFieldEff}
{\cal H} = - {\bf n \cdot h} - \frac{ \beta }{ 4S } [{\bf n \times h}]^2
+ \frac{ \beta^2 }{ 12S^2 }({\bf n \cdot h})[{\bf n \times h}]^2 
+ O\left( \frac{ \beta^3 }{ S^3 } \right),
\end{equation}  
where ${\bf h} \equiv S {\bf H}$ is the scaled magnetic field.
A similar quasiclassical Hamiltonian for a particle in a potential $V(x)$ has been
recently obtained with the help of cumulants  in ref.\ \cite{klaful98}.

The function ${\cal Z}$ can be evaluated from eq.\ (\ref{ZCohState}) as for a classical system
by doing the integral over the directions of the vector ${\bf n}$. 
Physical quantities such as the energy $U$ and magnetization ${\bf m}$ can be
obtained by differentiating ${\cal Z}$ with respect to the appropriate parameter.
Direct evaluation of observables as classical averages $\langle \ldots\rangle$ 
must be done with care.
For the energy one finds
%
\begin{equation}\label{HStar}
U = - \frac{ \partial \ln {\cal Z} }{ \partial \beta } = \langle {\cal H}^*\rangle,
\qquad 
{\cal H}^* = \frac{\partial (\beta{\cal H}) }{\partial \beta } 
= {\cal H} -  \frac{ \beta }{ 4S } [{\bf n \times h}]^2
+ O\left( \frac{ \beta^2 }{ S^2 } \right),
\end{equation}  
whereas the scaled magnetization ${\bf m}\equiv \langle {\bf S} \rangle/S$
 is given by
%
\begin{equation}\label{nStar}
{\bf m} = - \frac{ \partial \ln {\cal Z} }{ \partial (\beta{\bf h}) } 
= \langle {\bf n}^*\rangle,
\qquad 
{\bf n}^* = - \frac{\partial {\cal H} }{\partial {\bf h} } = {\bf n} 
-  \frac{ \beta }{ 2S } [{\bf n\times}[{\bf n \times h}]]
+ O\left( \frac{ \beta^2 }{ S^2 } \right).
\end{equation}  
One notices that not only the effective quasiclassical Hamiltonian ${\cal H}$ but
also the averaged quantities contain quantum corrections.

Next we have to find the limits of applicability of the quasiclassical expansion.
It is clear that the method will break down at low temperatures.
At such temperatures the spin fluctuates by a small angle $\theta$ from the
direction of the field, such that the deviation of the zeroth-order term of 
$\beta {\cal H}$ [see eq.\ (\ref{HamFieldEff})] from the ground state is 
$\beta h \theta^2 \sim 1$.
The second- and third-order terms of $\beta {\cal H}$ become 
$\beta h \theta^2 (\beta h/S)$ and $\beta h \theta^2 (\beta h/S)^2$, respectively.
Thus the quasiclassical expansion converges for $\beta h/S \lesssim 1$, i.e.,
for temperatures $T \gtrsim h/S$ which may be well below the classical energy
scale $h$ for large spins.
For comparison, a usual high-temperature series expansion converges for 
$T \gtrsim h$.
The quantum energy scale $h/S=H$ is defined by the separation of energy levels of the
spin in the field.
Replacement of discrete energy levels by a continuum of classical spin orientations
leads to a temperature dependence of the effective Hamiltonian ${\cal H}$ or of 
the density of states, if $\exp(-\beta {\cal H})$ in 
eq.\ (\ref{ZCohState}) is expanded in powers of $1/S$.
This expansion is justified under the same condition $\beta h/S \lesssim 1$.
For $\beta h/S \gg 1$, i.e., $T \ll h/S$, the spin remains in its ground state, and a continuous 
quasiclassical expansion becomes invalid.
The form of ${\cal H}$ in this region is obtained by setting $m=S$ in 
eq.\ (\ref{HamFieldHeffExact}): ${\cal H} = -SH -2ST\ln[(1+\cos\theta)/2]$.

\begin{figure}[t]
\unitlength1cm
\begin{picture}(7,6)
\epsfig{file=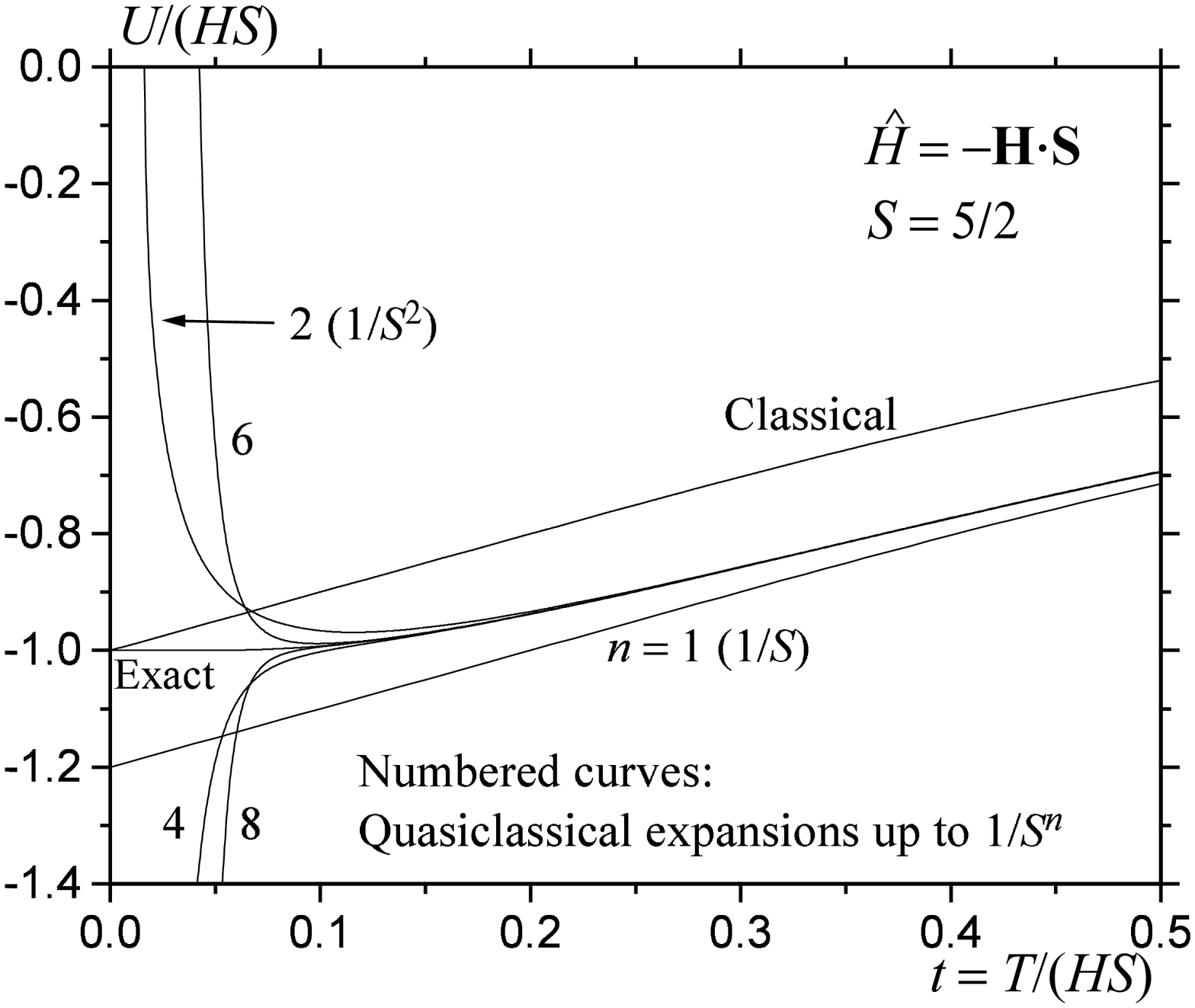,angle=-90,width=7.5cm}
\end{picture}
\begin{picture}(7,6)
\epsfig{file=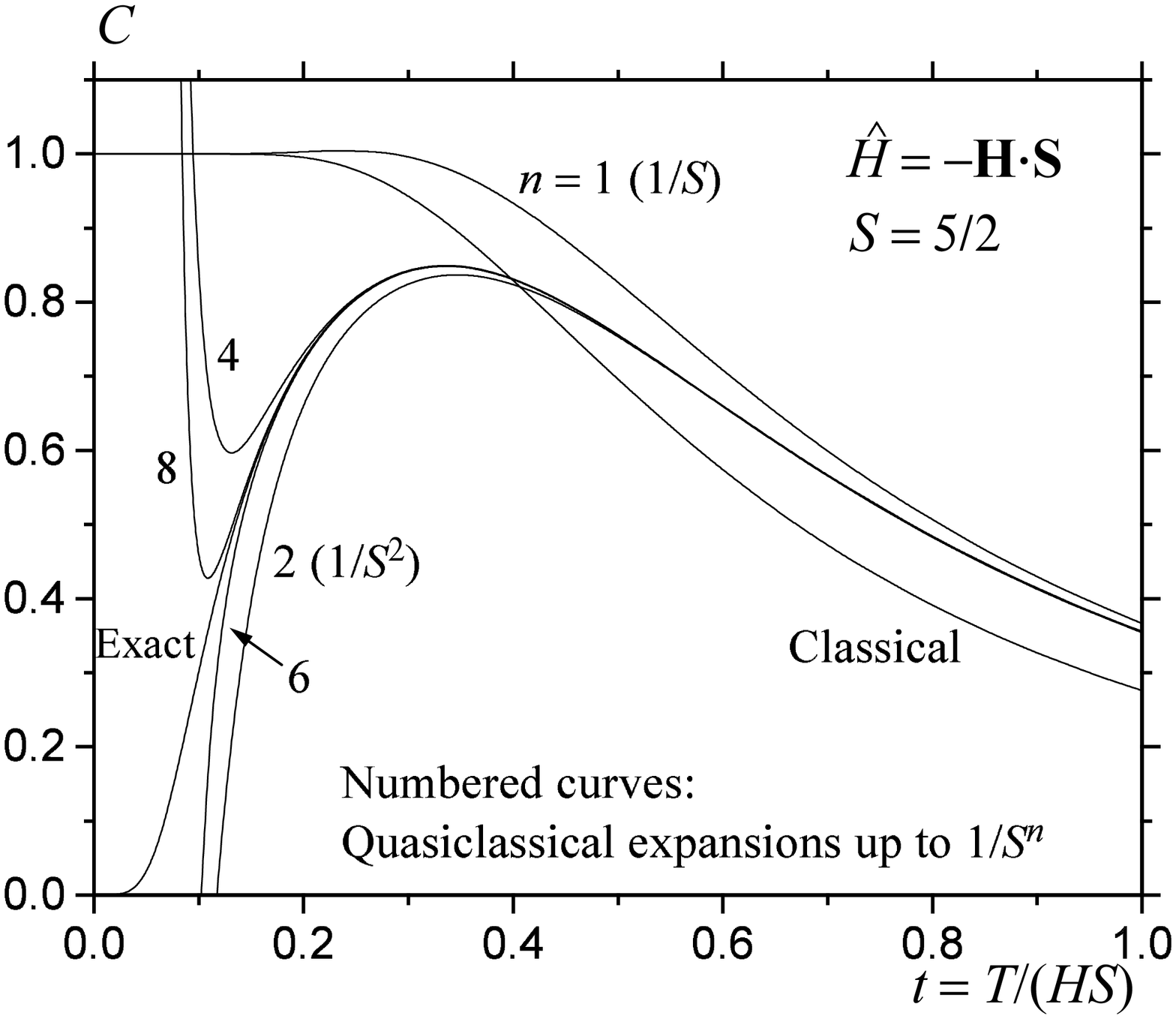,angle=-90,width=7.5cm}
\end{picture}
\caption{ \label{cum_h52}
Quasiclassical expansions of energy and heat capacity for the spin-in-a-field
model, eq.\ (\protect\ref{HamField}).
}
\end{figure}

\begin{figure}[t]
\unitlength1cm
\begin{picture}(7,6)
\epsfig{file=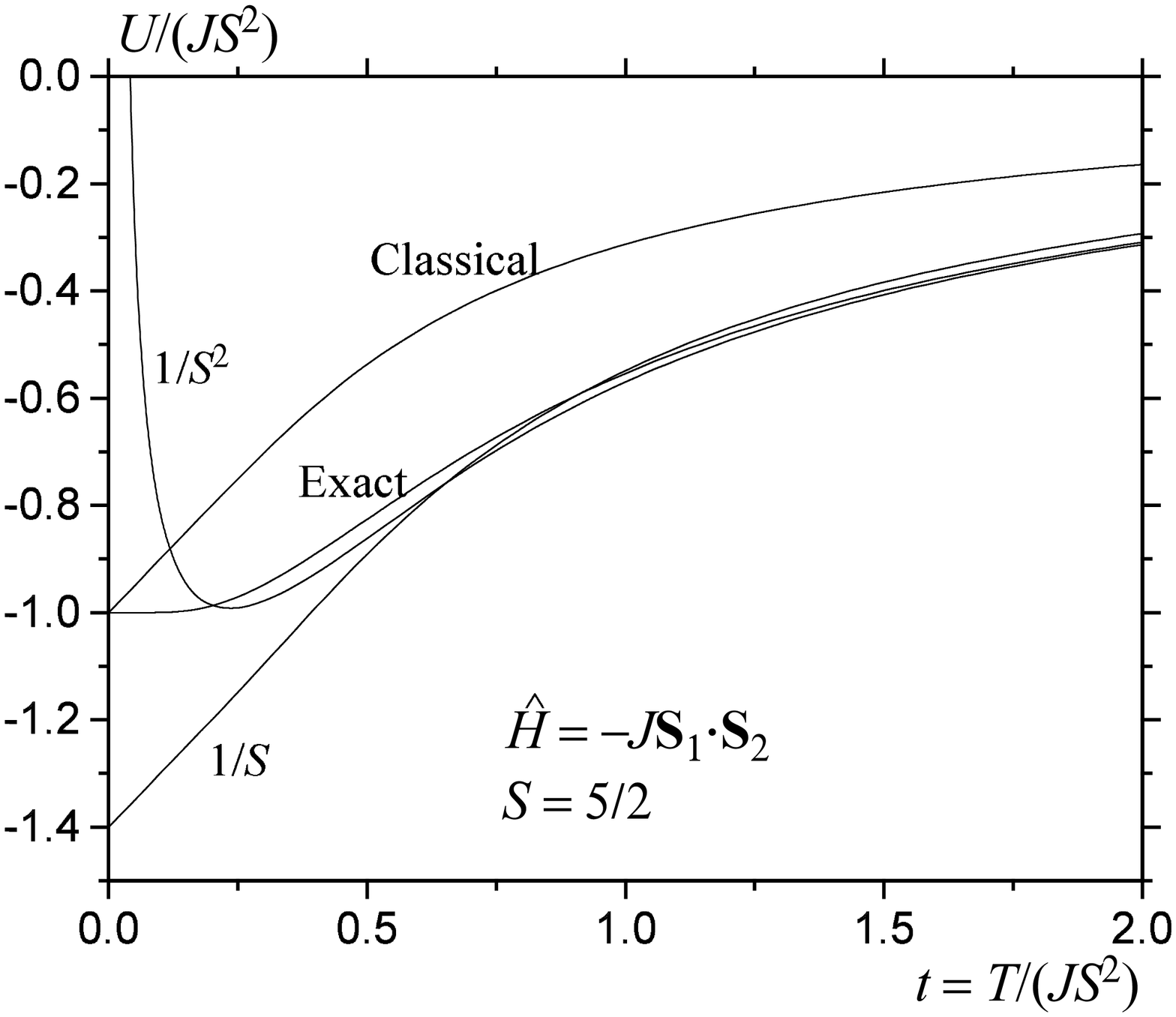,angle=-90,width=7.5cm}
\end{picture}
\begin{picture}(7,6)
\epsfig{file=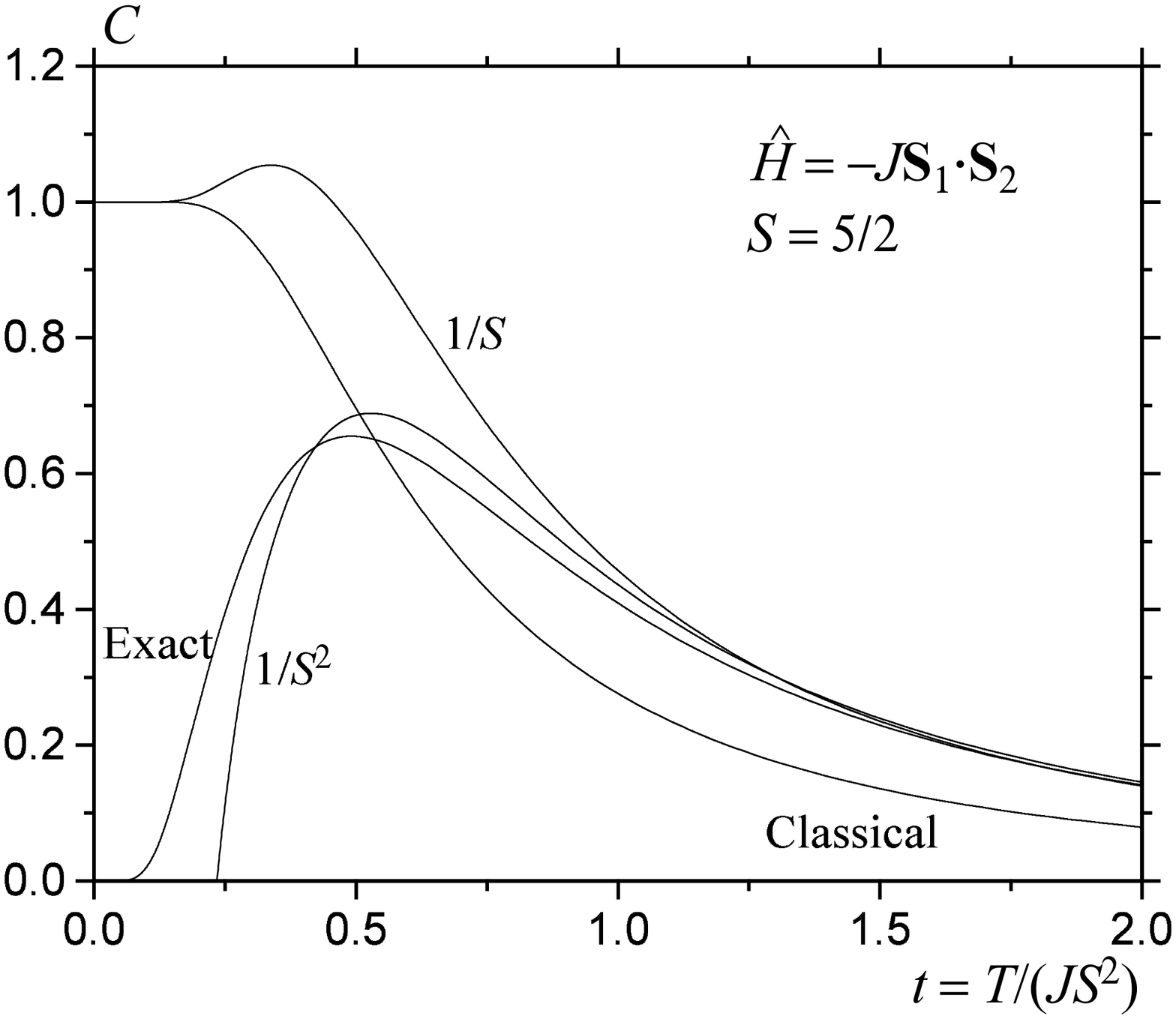,angle=-90,width=7.5cm}
\end{picture} \\
\begin{picture}(7,6)
\epsfig{file=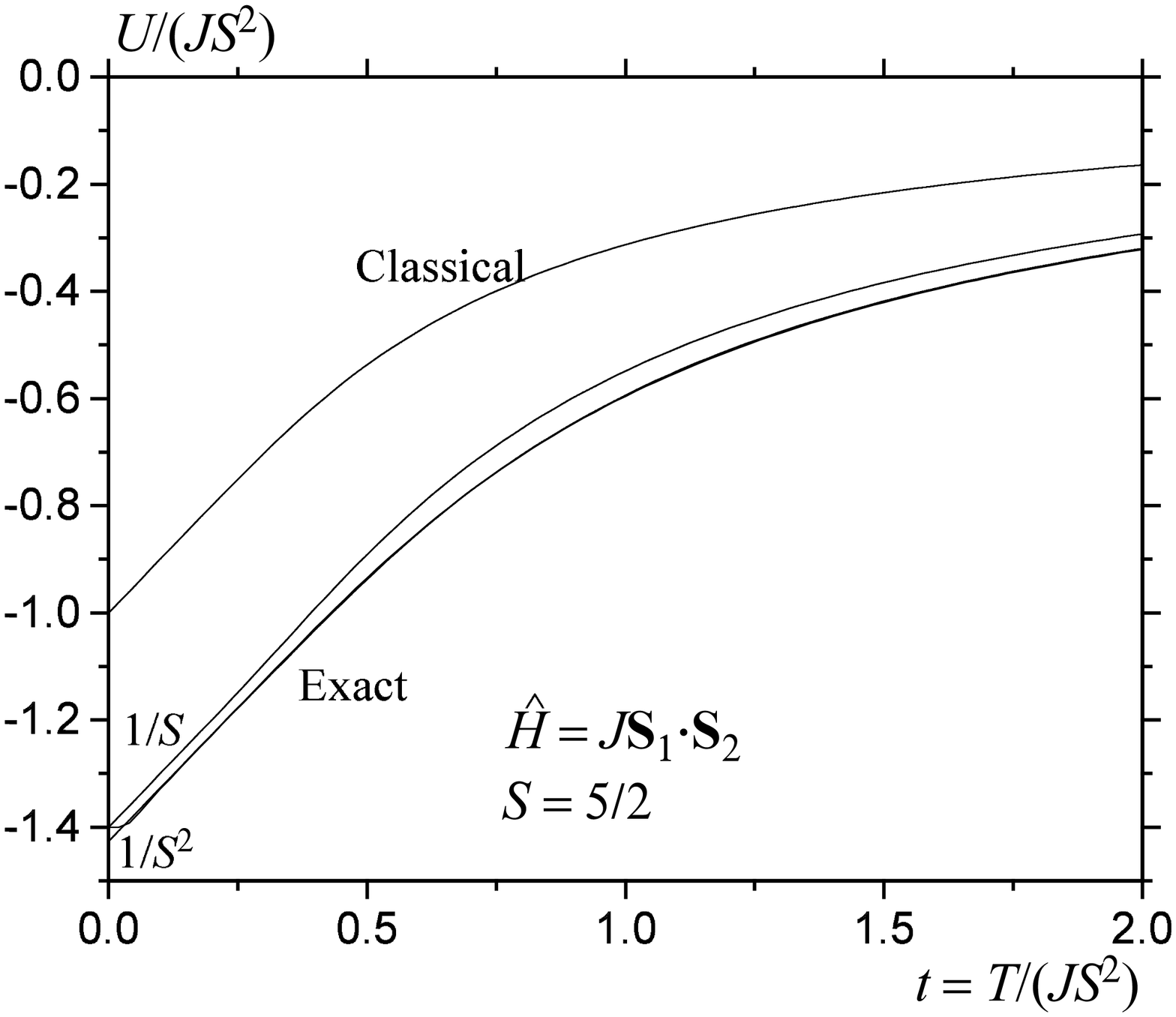,angle=-90,width=7.5cm}
%
\end{picture}
\begin{picture}(7,6)
\epsfig{file=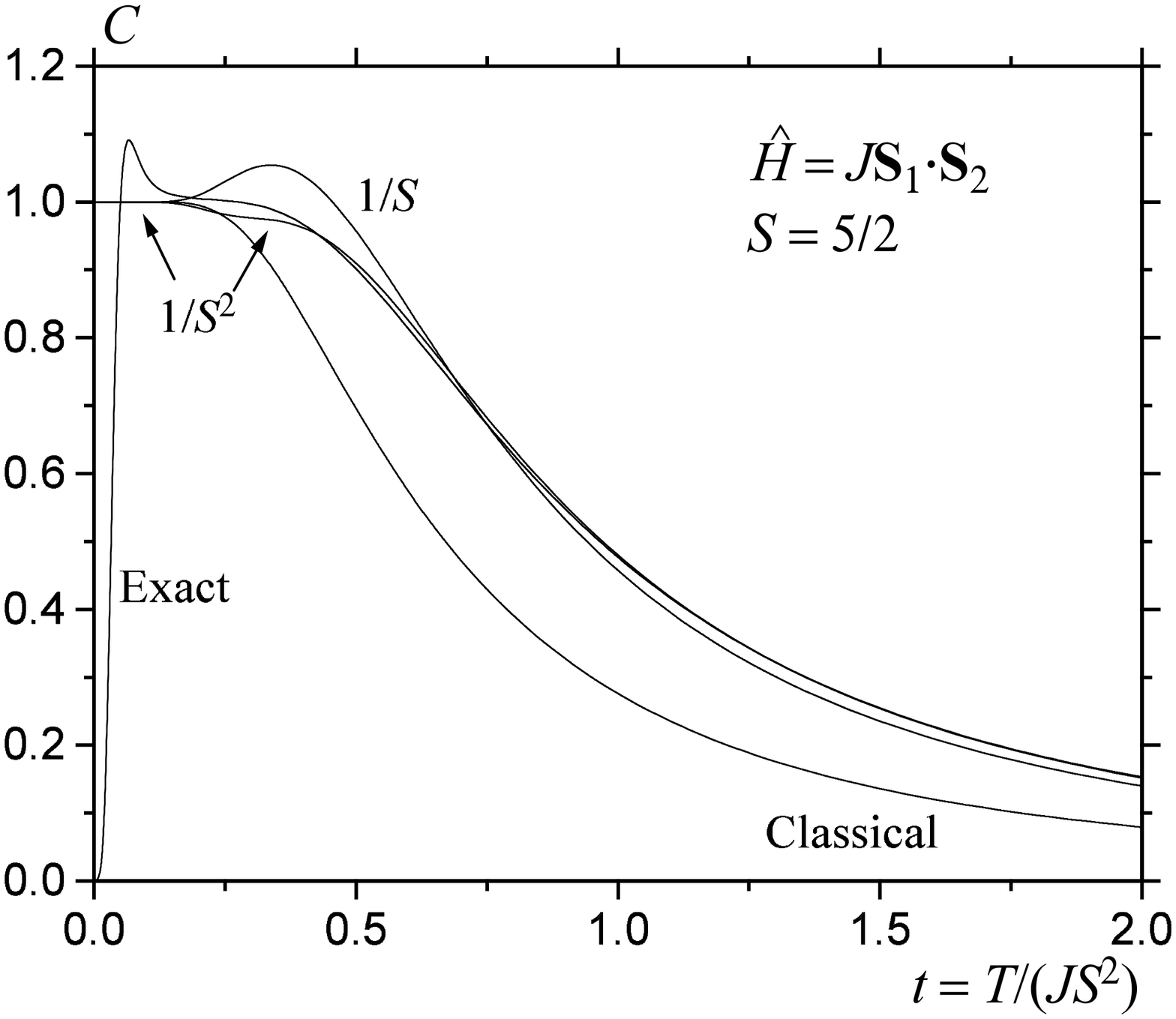,angle=-90,width=7.5cm}
\end{picture}
\caption{ \label{cum_tw52}
Quasiclassical expansion for the energy and heat capacity of the ferro- and
antiferromagnetic two-spin
models, eq.\ (\protect\ref{HamTwoSpins}).
For the antiferromagnetic coupling, the $1/S$ approximation reproduces the exact
ground-state energy, $U_0 = -1.4 \tilde J$, whereas the $1/S^2$ approximation
gives $U_0 = -1.442666 \tilde J$. 
}
\end{figure}

If $\exp(-\beta {\cal H})$ is expanded in powers of $1/S$, the breakdown of the quasiclassical 
expansion occurs at lower temperatures and the overall results are better. 
In this case an analytical calculation is possible.
Using eq.\ (\ref{HamFieldEff}),  one obtains to order $1/S^2$
%
\begin{equation}\label{LnZField}
\ln {\cal Z} \cong \ln(2S) + \ln \left( \frac{\sinh \xi }{ \xi } \right)
+ \frac{ \xi B + 1 }{2S} + \frac{ 2\xi^2 -3\xi^2B^2 -6\xi B - 3 }{ 24S^2 }
+ O\left(\frac{ 1}{ S^3 }\right),
\end{equation}  
where $\xi \equiv \beta h$ and $B=B(\xi)= \coth\xi - 1/\xi$ is the Langevin
function.
Quasiclassical expansions of different order in $1/S$ for the energy and  heat
capacity of the spin-in-a-field model are shown in fig. \ref{cum_h52}.
Expansions of orders higher than two have been obtained directly by expanding the
exact partition function ${\cal Z}=\sinh\{\xi[1+1/(2S)]\}/\sinh[\xi/(2S)]$ in powers of 
$1/S$.
It is seen that already the second quantum approximation nearly
quantitatively reproduces the maximum of the heat capacity, although the spin
value $S=5/2$ is not very high.

As a second example, let us consider the model of two ferro- or
antiferromagnetically coupled spins
%
\begin{equation}\label{HamTwoSpins}
\hat H = \mp J {\bf S}_1 \cdot  {\bf S}_2.
\end{equation}  
This model is exactly solvable.
The eigenstates labelled by the total spin 
$n=0, 1, \ldots, 2S$ have the energy $E_n = \mp J [n(n+1)/2 - S(S+1)]$ and
degeneracy $2n+1$.
The ferro- and antiferromagnetic ground states are $E_{2S}=-\tilde J$ and
$E_0=-\tilde J(1+1/S)$, respectively, where $\tilde J \equiv JS^2$ is the
classical energy scale.
The separation of levels near the ground state is $\Delta E \sim JS =\tilde J/S$
for the ferro- and $\Delta E \sim J =\tilde J/S^2$ for the antiferromagnet.

Derivation of the effective Hamiltonian ${\cal H}({\bf n}_1, {\bf n}_2)$ for this
model, where ${\bf n}_{1,2}$ are the coherent-state vectors at both sites, proceeds
along the same lines.
Eqs.\ (\ref{ZCohState})--(\ref{HEffExp}) retain their form, only the matrix elements
are defined with respect to both coherent states ${\bf n}_1$ and ${\bf n}_2$.
The latter slightly modifies the routine of calculating the cumulants of the
Hamiltonian $\hat H$: In eq.\ (\ref{HEffExp}) cumulants of $\hat H$ are rewritten
explicitly according to eq.\ (\ref{AvrToCum}), then independent averages on sites 1
and 2 are taken.
After that spins ${\bf S}_1$ and ${\bf S}_2$ are expressed in their own coherent-state
coordinate systems and the matrix elements 
of their components $S_{1\alpha_1}$ and $S_{2\alpha_2}$ are computed with the help of cumulants.
In a cumulant expansion of eq. (\ref{HEffExp}),
this modification results in the
appearance of terms carrying higher 
powers of $1/S$, in addition to the leading terms of order $(\beta/S)^n$.
To order $1/S^2$ one obtains
%
\begin{eqnarray}\label{HamEffTwoSpins}
&&
{\cal H} = \mp \tilde J {\bf n}_1 \cdot  {\bf n}_2
- \frac{ \beta \tilde J^2 }{ 2S } [1 - ({\bf n}_1 \cdot  {\bf n}_2)^2]
- \frac{ \beta \tilde J^2 }{ 8S^2 } [1 - {\bf n}_1 \cdot  {\bf n}_2]^2
\nonumber\\
&&\qquad
{} \mp \frac{ \beta^2 \tilde J^3 }{ 12S^2 }
[1 - ({\bf n}_1 \cdot  {\bf n}_2)^2] [1 - 5 {\bf n}_1 \cdot  {\bf n}_2]
+ O\left(\frac{ 1}{ S^3 }\right).
\end{eqnarray}  
Note that quantum corrections have a non-Heisenberg form.
As follows from the cumulant expansion of eq.\ (\ref{HEffExp}),
an additional pair ${\bf n}_1{\bf \cdot n}_2$ is associated with each further 
power of $\beta$.
Eq.\ (\ref{HamEffTwoSpins}) results in the expansion for $\ln {\cal Z}$ of the form 
%
\begin{equation}\label{LnZTwoSpins}
\ln {\cal Z} \cong 2\ln(2S) + \ln \left( \frac{ \sinh \xi }{ \xi } \right)
+ \frac{ \xi B + 1 }{S} + \frac{ 5\xi^2 -6\xi^2B^2 \mp \xi^2 B -9\xi B - 3}{ 12S^2 }
+ O\left(\frac{ 1}{ S^3 }\right),
\end{equation}  
where $\xi\equiv \beta\tilde J$.

Quasiclassical expansions for the energy and  heat
capacity of the two-spin model are shown in fig.\ \ref{cum_tw52}.
They are obtained by differentiating eq.\ (\ref{LnZTwoSpins}).
For ferromagnetic coupling the results resemble those for the spin-in-a-field
model, although here quantum corrections are about two times larger.
For that model, the leading asymptote $U \cong -h \xi (1+1/S)/3$  at high temperatures
 is recovered already at order  $1/S$, whereas
the proper leading asymptote $U \cong -\tilde J \xi (1+1/S)^2/3$ for 
the two-spin model is obtained only at order $1/S^2$.

For antiferromagnetic coupling and to order $1/S^2$, $U$ and $C=\drm U/\drm T$  do
not diverge as $T\to 0$, because the leading $\xi^2$ terms in 
eq.\ (\ref{LnZTwoSpins}) cancel.
The accuracy of the $1/S^2$ approximation over a wide range of temperatures
 is quite impressive.
It is tempting to explain this by the fact that for  antiferromagnetic
coupling the separation of the low-lying energy levels is by a
factor of $1/S \ll 1$ smaller than for ferromagnetic coupling, as was mentioned 
above.
Therefore the properties of the system are expected to be ``more classical''.

Furthermore, one could speculate that for many-spin systems without an energy gap
the quasiclassical expansion works even better.
However, it is not true.
The quasiclassical 
expansion without a resummation does not reproduce {\em rational}
powers of $T$, which appear, e.g., in linear spin-wave theory.
Therefore, it should break down below the quantum energy scale $JS=\tilde J/S$.
Note that the SWT based either on the Holstein-Primakoff or
path-integral formalisms (see, e.g., ref.\ \cite{aue94book}) works in the range
$T \lesssim \tilde J$.
Its results should overlap with those of the cumulant
expansion in the range $\tilde J/S \lesssim T \lesssim \tilde J$.    
It is interesting to speculate how $\beta/S$ corrections of the
cumulant expansion go over to $1/S$ corrections of the SWT (e.g., for
antiferromagnets) when $T \lesssim \tilde J/S$.
One would expect that the leading terms $(\beta/S)^n$ of
eq.\ (\ref{HamEffTwoSpins}) convert to the leading order of the SWT, whereas
 $\beta/S^2$ and other  subdominant terms 
 result in  $1/S$ spin-wave corrections.

We have shown that for spin systems with large $S$ an application of the cumulant technique 
together with spin coherent states leads to effective, classical-like
Hamiltonians with quantum corrections taken into account.
These Hamiltonians can be further treated by a number of methods for
classical spin systems, such as exact solutions, Monte-Carlo simulations, or the
classical-spin diagram technique \cite{gar96prb}, which is again based on
cumulants.

%

\end{document}